\documentclass[letterpaper, 10 pt, conference]{ieeeconf}  

\usepackage{cite}
\usepackage{float}
\usepackage{multicol}
\usepackage{multirow}
\usepackage{hyperref}
\usepackage{graphicx}
\usepackage{amsfonts}
\usepackage{amsmath}
\usepackage{color}
\usepackage{epstopdf}
\usepackage{subcaption}
\usepackage{multirow}
\usepackage{bbm}
\usepackage{bm}
\usepackage{comment}

\newcommand{\D}{\mathrm d}
\newcommand{\pD}{\partial }

\title{\LARGE \bf
Adiabatic approach for natural gas pipeline computations
}

\author{ \parbox{2.4 in}{\centering {\bf Michael Chertkov} \\
        Theory Division, T-4  \& CNLS\\ of
        Los Alamos National Laboratory,\\ Los Alamos, NM 87544, USA;\\
        \& Skolkovo Institute
        of Science and Technology,
        Moscow, Russia}
         \parbox{2.4 in}{ \centering {\bf Alexander Korotkevich} \\
         Department of Math \& Statistics\\ of University of New Mexico,\\
         Albuquerque, NM 87131, USA;\\
         \& L.D. Landau Institute for Theoretical Physics,
         Moscow, Russia}
}

\begin{document}

\maketitle
\thispagestyle{empty}
\pagestyle{empty}

\begin{abstract}
We consider slowly evolving, i.e. ADIABATIC, operational regime within a transmission level (continental scale) natural gas pipeline system.  This allows us to introduce a set of nodal equations of reduced complexity describing  gas transients in  injection/consumption UNBALANCED (so-called line-pack) cases. We discuss, in details, construction of the UNBALANCED ADIABATIC (UA) approximation on the basic example of a single pipe. The UA approximation is expected to play a significant ``model reduction" role in solving control, optimization and planning problems relevant for flawless functioning of modern natural gas networks.
\end{abstract}

\vspace{-0.3cm}
\section{Introduction}

Natural gas transmission pipes extend over continents. It is important to plan, build and operate flow of gas through the system, which is injected at the gas terminals or reservoirs, compressed at the compressor stations and withdrawn/consumed by industrial and private customers.   Coordinating seamless work of all the components of the system poses a significant challenge. Modeling and simulations of the gas flow through the system is critical for all problems in the system design, optimization and control. In the normal operational regime considered in this manuscript (no fast/abrupt changes of an emergency type) starting point for modeling gas flow is a system of spatio-temporal and nonlinear PDEs, two per-pipe describing mass and momentum transfer, supplemented by boundary conditions at the junctions and compressors, and thermodynamic relations describing pressure and flow transformations at various nodal elements of the systems \cite{71Suk,osiadacz84,thorley87,00SK,Sardanashvili2005}. For majority of operational problems (as well as many planning problems) this large system of equations need to be solved in the most general dynamic case \cite{TT1987,ZA2000,BHK2006,banda08,HMS2010}, where consumption and production of gas are not balanced. This unbalanced regime is characterized by the so-called line-pack, thus emphasizing that gas can accumulate or be withdrawn in different parts of the systems globally unbalanced thus leading to significant pressure transients.

Solving the system of nonlinear Partial Differential Equations (PDEs) in direct simulations is prohibitively expensive. To overcome the challenge, researchers have focused on looking for approximate but efficient computational techniques, see e.g. \cite{chua82,osiadacz84,thorley87,kralik88,Osiadacz1989,Kiuchi1994,hudson06}. In particular, of a great interest are the so-called lump-element computational methods \cite{75SS,HMS2010,CBL2015,15SP,grundel14,15ZCB,15ZDBC} which allow to replace the PDE model by an Ordinary Differential Equations (ODEs) model, where the ODEs are describing nonlinear dynamics of relevant parameters (pressure, mass flow, and temperature) only at a limited number of spatial positions along the system. Moreover, an ultimate goal in  this PDE-to-ODE reduction consists in keeping in the description only critical nodes, such as end-nodes of the pipes, compressor locations, branching points and injection/consumption locations \cite{75SS,HMS2010,CBL2015,15SP}. Two flavors of the the lump-element methods were reported in the literature. First, it was suggested to approximate spatial derivatives through a properly chosen discrete approximation, e.g. resembling finite-element schemes of different types \cite{75SS,15SP,grundel14,15ZCB,15ZDBC}. This approach would normally be validated empirically through comparison with high resolution methods (which are less efficient, but more accurate). On the contrary, this manuscript focuses on developing lump-element models of the second, so-called adiabatic, type \cite{HMS2010,CBL2015}. The adiabatic models are exact in the asymptotic regimes  where characteristic time scale of the input parameters (e.g. injection/consumption) becomes infinite. (Some other restrictions on an adiabatic method validity may apply as well -- see below.) When the input time scale is large but finite,  the adiabatic methods provide leading results with corrections, which can be computed (at least in principle) systematically. (For the method to work well the corrections should be sufficiently small with respect to the small adiabatic parameter stated as the ratio of the largest natural time scale of the system to the exogenous time scale of the slowly changing input.) First adiabatic approximation was suggested in \cite{HMS2010}, where it was noticed that allowing parameters of a stationary flow solution to evolve slowly allows to represent at least some part of the actual solution temporal dynamics.  The approximation was improved in \cite{CBL2015} where it was shown how to make the adiabatic description self-consistent.  Systematic (as exact in the limit when input/output characteristics freeze) and self-consistent approximation was derived for a gas network in \cite{CBL2015}. It was shown in \cite{CBL2015} how the approximation results in the reduced description relating dynamics of pressures and mass flows to the exogenously fixed injections/consumption or pressures at the critical nodes only. However, one significant handicap of the adiabatic approximation of \cite{CBL2015} consisted in the fact that the approximation also required, in addition to the slowness of the inputs, that the transients remain sufficiently close to possibly different but still balanced solutions. Smallness of the deviations explored in \cite{CBL2015} was critically linked to existence of a class of stationary, i.e. time independent and thus balanced, solutions of the gas flow equations which are analytically tractable. However, when deviations from the balance situation persists for sufficiently long time the Balanced Adiabatic (BA) approximation of \cite{CBL2015} does not apply.

In this manuscript we address this caveat of the BA approximation and show how to generalize it and thus construct Unbalanced Adiabatic (UA) solution which is not limited to the balanced case. Our approach consists of the following two steps:
\begin{itemize}
\item Construction of a novel family of exact unbalanced, and thus non-stationary, solutions of the basic gas-flow PDEs. The non-stationary solutions grow or decay exponentially in time, where thus the time-independent (stationary) and balanced case being a marginal example.
\item Development of the adiabatic generalization of the exact unbalanced solutions allowing for slow evolution of the unbalanced solution parameters. This new UA construction generalizes the BA construction described in \cite{CBL2015}, thus allowing for significant deviations from the balanced (stationary) case. 
\end{itemize}
Even though the construction applies to networks,  we focus in this manuscript on the basic case of a single pipe,  thus leaving discussion of the network case for future publications.

The layout of material in the remaining part of the manuscript is as follows.
We introduce basic gas flow physics and modeling assumptions/considerations/equations in Section \ref{sec:TechIntro}. New family of exact dynamic solutions of the basic system of PDE for a single pipe is described in Section \ref{sec:exact}. Section \ref{sec:adiabatic_theory} details construction of the family of the UA solutions. Our theoretical construction is tested and validated against direct numerical simulations in Section \ref{sec:numerics}. Conclusions and path forward are discussed in Section \ref{sec:consutions}. Appendices are reserved for auxiliary/supplemental materials.

\vspace{-0.3cm}
\section{Modified Euler Equations for a Pipe}
\label{sec:TechIntro}

Natural gas is transmitted long-distances though a system of pipes in the turbulent, $Re\approx 10^6$, compressible/sub-sonic, $Ma\approx 1/30$ regime. Typical pressures in a $d=1$ m diameter pipe transmitting gas over hundreds to thousands of kilometers, is in the range of $p=300-800$ psi$\approx 21-55$ bar=$2.1-5.5 \times 10^6 kg/(m s^2)$, the gas flows with the speed of $u=10-15m/s$ which is significantly smaller that the speed of sound, $c_s\approx 300 m/s$. So-called Weymouth equations \cite{Osiadacz1989} accounting for dynamical balance of mass and momentum through a pipe segment of length $L$ (free of injection, consumption or compression) are \cite{wylie78,thorley87,hudson06}
 \begin{eqnarray}
 \label{mass_balance}&&
 \partial_t\rho+\partial_x (\rho u)=0,\\  \label{momentum_balance}
 && \partial_t (\rho u)+\partial_x (\rho u^2)+\partial_x p
 =-\frac{f\rho u |u|}{2d}
 \end{eqnarray}
where $x\in [0,L]$, $\rho$ is the mass density and velocity $u$ along the pipe is averaged over diameter of the pipe.
(Note that velocity profile resolved across the pipe is close to the equilibrium distribution where the average velocity has an almost flat profile everywhere except of a narrow boundary layer region near the walls.)
The natural gas is modeled as an ideal gas, thus pressure and density are in the standard thermodynamic relation,
$p=c_s^2 \rho$. Without loss of generality (generalizations are straightforward) we assume iso-thermal flow (temperature is constant) and ignore effects of gravity (tilted pipe). We also assume that the turbulent friction term in the momentum Eq.~(\ref{momentum_balance}) is (approximately) $Re$ number independent.
This approximation of the momentum equation applies in the regime where the friction term and the pressure terms are roughly in balance, while the dynamic (first) and self-advection (second) terms are respectively order and two orders of magnitude  smaller \cite{Osiadacz1989,HMS2010}. Indeed for the actual parameters of the transmission system mentioned above, one gets the following estimations for the ratio of the first to third terms, $|p\partial_t(1/\phi)|/\alpha\approx 0.1$, where $\alpha= f c_s^2/d\approx 900 m/s^2$ and one estimates temporal scale of the consumption/production variation in $100 s$,  and $\phi c_s/p\approx 0.03$ for the ratio of second to first terms. This means, in particular,  that description of this manuscript applies only to normal operations of the pipes when the dynamical changes take place on the scale of minutes or slower. The dynamics is typically driven by unsteady (but typical in terms of operations) changes of injection, consumption or compression. (In other words our description applies only to the regime when changes are forced by exogenous changes and no fast, high frequency, waves or shocks are present.)

In summary, under realistic assumption that the following dimensionless parameters are asymptotically small, $1\gg |p\partial_t(1/\phi)|/\alpha,|\phi| c_s/p$, a single-pipe dynamics is well approximated by the following system of equations
\begin{eqnarray}
\label{mass_balance_1} &&
c_s^{-2}\partial_t p+\partial_x \phi=0,\\
\label{momentum_balance_reduced} &&
\partial_xp+\alpha \frac{\phi|\phi|}{2p}=0.
 \end{eqnarray}
where the first equation is a version of Eq.~(\ref{mass_balance}) and the second equation  is the reduced version of Eq.~(\ref{momentum_balance}), both stated in terms of $\phi$ and $p$.

Note (for the sake of generality) that the formal singularity of Eqs.~(\ref{momentum_balance},\ref{momentum_balance_reduced}) at $v,\phi\to 0$ is not physical, as the conditions of the phenomenological derivation of the turbulence dissipation is obviously broken when the flow is slow, that is in the laminar (not turbulent) regime. In the laminar regime the dissipation is linear in velocity/flux. However, given that the regions of the small/laminar flux velocity, where the flow reverses its direction, are expected to be small in size (of the order of $d$ - the pipe diameter, we are ignoring this nuance in what follows.

\vspace{-0.3cm}
\section{Exact Unbalanced Solutions}
\label{sec:exact}

We look for solution of Eqs.~(\ref{mass_balance},\ref{momentum_balance_reduced}) in the following form
\begin{eqnarray}
\label{p_pipe} && p(t,x)=p_0 \exp\left(\frac{\lambda c_s^2}{\sqrt{2\alpha}} t+\psi_\lambda(x)\right),
\end{eqnarray}
where $p(t,x)\geq 0$.
Eq.~(\ref{p_pipe}) implies according to Eq.~(\ref{momentum_balance_reduced})
\begin{eqnarray}
\label{phi_2_pipe} && \phi(t,x)|\phi(t,x)|= \\ && \nonumber -\frac{2p_0^2}{\alpha}\exp\left(2\frac{\lambda c_s^2}{\sqrt{2\alpha}} t+2\psi_\lambda(x)\right) \psi_\lambda'(x).
\end{eqnarray}
Assume that $\forall t,x:\quad \phi>0$ (this assumption can be relaxed), then consistently with Eq.~(\ref{phi_2_pipe}), one finds that, $\forall x:\quad \psi_\lambda'(x)<0$, and substituting it into Eq.~(\ref{phi_pipe}) one arrives at
\begin{eqnarray}
\label{phi_pipe} && \phi(t,x)=\sqrt{-\frac{2p_0^2}{\alpha}\psi_\lambda'(x)}\exp\left(\frac{\lambda c_s^2}{\sqrt{2\alpha}} t+\psi_\lambda(x)\right).
\end{eqnarray}
Substituting Eqs.~(\ref{p_pipe},\ref{phi_pipe}) into  Eqs.~(\ref{mass_balance_1}) one finds that $\psi_\lambda(x)$ satisfies,
\begin{eqnarray}
\lambda \sqrt{-\psi_\lambda'}-2(\psi_\lambda')^2-\psi_\lambda''=0.\label{psi_eq}
\end{eqnarray}
Denoting, $G(x;\lambda)=-\psi_\lambda'$, and then,
$\psi_\lambda(x)=-\int_0^x dx' G(x';\lambda)$,
one gets from Eq.~(\ref{psi_eq}) that $G$ satisfies
\begin{eqnarray}
\lambda \sqrt{G}-2 G^2+G'=0.
\label{G-eq}
\end{eqnarray}
Integration of this equations results in the following  formula expressing, $G$, via $x$, $\lambda$ and $G_0$ implicitly
\begin{eqnarray}
\int\limits^{G_0}_{G(x;\lambda, G_0)} \frac{dz}{\lambda\sqrt{z}-2z^2}=x. \label{G_implicit}
\end{eqnarray}
The integral on the lhs of Eq.~(\ref{G_implicit}) can be computed in quadratures.
Introduce the following function:
\begin{eqnarray}
&& f(z,\lambda) = \frac{2^{2/3}}{3\lambda^{2/3}}\Biggl(-\log\left(\frac{\lambda^{1/3}}{2^{1/3}}-\sqrt{z}\right) \nonumber\\
&&  + \frac{1}{2}\log\left(\frac{\lambda^{2/3}}{2^{2/3}} + \frac{\lambda^{1/3}}{2^{1/3}}\sqrt{z} + z\right) \nonumber\\
&& + \sqrt{3}\arctan\left(\frac{1}{\sqrt{3}}\left(1+2\sqrt{z}\frac{2^{1/3}}{\lambda^{1/3}}\right)\right)\Biggr),\nonumber 
\end{eqnarray}
then ~\eqref{G_implicit} becomes,
$f(G_0,\lambda) - f(G,\lambda) = x$. 

It is instructive to check the special case of $\lambda=0$. Then by direct integration of~\eqref{G_implicit} one arrives at
\begin{equation}
G(x,0,G_0) = \frac{G_0}{1 - 2 G_0 x},\nonumber 
\end{equation}
obviously corresponds to the stationary solution, $p=p_0\sqrt{1-2 G_0 x}$.

\vspace{-0.3cm}
\section{Unbalanced Adiabatic Solutions: Theory}
\label{sec:adiabatic_theory}

In this Section we explain how to generalize the Balanced-Adiabatic (BA) approach of \cite{CBL2015}, developed based on the balanced/stationary solution of Eqs.~(\ref{mass_balance},\ref{momentum_balance_reduced}), to a new Unbalanced-Adiabatic (UA) based on  the exact solutions described in the previous Section. The idea of how we extend the exact solution into an adiabatic one remains the same:  we allow two parameters describing the exact solution to have in addition dependence on time.  The dependence is assumed slow. Formally, it means that time derivatives of parameters depending on time are sufficiently small (when compared with other time scales of the system and of the base solution), but not the parameters themselves. We substitute adiabatic expressions for pressure and mass flow with added corrections of the general type into Eqs.~(\ref{mass_balance},\ref{momentum_balance_reduced}), linearize over temporal derivatives of the slow parameters and unstructured small corrections to establish self-consistent relations between the terms.  These relations will then allow us to link temporal evolution of the two adiabatic parameters and two conditions for pressures or mass flows at the ends of the pipe. (Different boundary conditions correspond to different problems of interest.)

Consider the case when the boundary conditions in the dynamic solutions explained above in Section \ref{sec:exact} change slowly in comparison with the original/bare dynamics. Specifically,  let us assume that the boundary conditions for pressure on two ends of the pipe are parameterized by  $\lambda(t)$ and $G_0(t)$ according to
\begin{eqnarray}
\label{p_boundary0} && \hspace{-1cm} p(t,0)=p_0 \exp\left(\frac{c_s^2}{\sqrt{2\alpha}} \int_0^tdt' \lambda(t')\right),
\\ && \hspace{-1cm}
p(t,L)=p(t,0)\exp\left(-\int\limits_{0}^{L} dx' G(x';\lambda(t), G_0(t))\right),
\label{p_boundaryL}
\end{eqnarray}
where the dependence of $G(x;\lambda, G_0)$ on $x$ is set implicitly by Eqs.~(\ref{G_implicit}).

Let us look for approximate solution of Eqs.~(\ref{mass_balance_1},\ref{momentum_balance_reduced}) in the form
\begin{eqnarray}
&& p(t;x)=p_{\mbox{\small UA}}(x;\lambda(t),G_0(t))+\delta p(t;x)\label{adiabatic_p}\\
&& \phi(t;x)=\phi_{\mbox{\small UA}}(x;\lambda(t),G_0(t))+\delta \phi(t;x). \label{adiabatic_phi}
\end{eqnarray}
where
\begin{eqnarray}
\label{p-base}
&& p_{\mbox{\small UA}}(x;\lambda(t),G_0(t))= \\
&& p_0\exp\left(\frac{c_s^2}{\sqrt{2\alpha}}\int\limits_0^t dt' \lambda(t')\!-\!\!
\int\limits_{0}^{x} dx' G(x';\lambda(t),G_0(t))
\right), \nonumber\\ \label{phi-base}
&& \phi_{\mbox{\small UA}}(x;\lambda(t),G_0(t)) = \\ && \sqrt{\frac{2}{\alpha}} p_{\mbox{\small UA}}(x;\lambda(t), G_0(t))\sqrt{G(x;\lambda(t),G_0(t))},
\nonumber
\end{eqnarray}
is the adiabatic (slowly changing) part of the solution and $\delta p(t;x)$ and $\delta \phi(t;x)$ are respective perturbative corrections. (We also assume here that the dynamics does not drive a flow reversal within the pipe. Generalization is straightforward.)
Then, substituting Eqs.~(\ref{adiabatic_p},\ref{adiabatic_phi},\ref{p-base},\ref{phi-base}) into Eqs.~(\ref{mass_balance_1},\ref{momentum_balance_reduced}) and  linearizing the result over $\delta p$, $\delta\phi$ one arrives at
\begin{eqnarray}
\nonumber && \partial_x \delta \phi(t;x) +c_s^{-2}\partial_t \delta p(t;x)
+\partial_x \phi_{\mbox{\small UA}}(x;\lambda(t),G_0(t))\\
&& +c_s^{-2}\partial_t p_{\mbox{\small UA}}(x;\lambda(t),G_0(t))=0
,\label{mass-balance-adiabatic}\\
\label{momentum-balance-adiabatic} &&
\partial_x \left(\delta p(t;x) p_{\mbox{\small UA}}(x;\lambda(t),G_0(t))\right)\nonumber\\ && +\alpha \delta \phi(t;x) \phi_{\mbox{\small UA}}(x;\lambda(t),G_0(t))=0.
\end{eqnarray}
Integrating Eq.~(\ref{momentum-balance-adiabatic}) over the entire pipe and taking into account that
$\delta p(t;0)=\delta p(t;L)=0$ (correspondent to the boundary conditions exogenously fixed) one derives
\begin{equation}
\int\limits_0^L dx \delta \phi(t;x) \phi_{\mbox{\small UA}}(x;\lambda(t),G_0(t))=0.
\label{momentum-balance-adiabatic-1}
\end{equation}
On the other hand the adiabatic approximation suggests to ignore the second term in Eq.~(\ref{mass-balance-adiabatic}) in comparison with the other terms. Integration of Eq.~(\ref{mass-balance-adiabatic}) with the second term droped results in
\begin{eqnarray}
\nonumber
&&\delta \phi(t;x)=-\int\limits_0^x dx'
\Biggl( \partial_{x'} \phi_{\mbox{\small UA}}(x';\lambda(t),G_0(t)) \\ && +c_s^{-2}\partial_t p_{\mbox{\small UA}}(x';\lambda(t),G_0(t))\Biggr)
+\delta \phi(t;0).
\label{mass-balance-adiabatic-1}
\end{eqnarray}
Taking into account Eqs.~(\ref{momentum-balance-adiabatic-1}) one derives
\begin{eqnarray}
\label{delta_phi_0}
&& \hspace{-1cm}\delta\phi(t;0)=\int\limits_0^L \frac{dx \phi_{\mbox{\small UA}}(x;\lambda(t),G_0(t))}{\int\limits_0^L dx \phi_{\mbox{\small UA}}(x;\lambda(t),G_0(t))}
\int\limits_0^x dx'  \\ && \hspace{-1cm}
\left( \partial_{x'} \phi_{\mbox{\small UA}}(x';\lambda(t),G_0(t))+c_s^{-2}\partial_t p_{\mbox{\small UA}}(x';\lambda(t),G_0(t))\right).
\nonumber
\end{eqnarray}
Further, respective expression for $\delta p(t;x)$ can be obtained from~\eqref{momentum-balance-adiabatic} integrating
both sides of the equation from $0$ to $x$, thus accounting for the boundary condition $\delta p(t;0)=0$:
\begin{equation}
\delta p(t;x) = -\frac{\alpha\int\limits_{0}^{x}\D x\phi_{\mbox{\small UA}}(x;\lambda(t),G_0(t)) \delta \phi(t;x)}{p_{\mbox{\small UA}}(x;\lambda(t),G_0(t))}. \label{delta_p}
\end{equation}

In summary, Eqs.~(\ref{adiabatic_p}-\ref{delta_p}) provide explicit dependence of pressure and mass flow within the pipe on only two time-dependent parameters, $\lambda(t)$ and $G_0(t)$.  The two parameters can be expressed via two boundary conditions at the two ends of the pipe,  e.g. two pressures at the inlet and outlet or two flows or a mix (say pressure at inlet and flow at the outlet). Relations between the parameters (also involving according to the equations above their temporal derivatives) result in the system of ODEs  which replace the original description based on PDEs~(\ref{mass_balance},\ref{momentum_balance_reduced}) we have started from.

An alternative choice of boundary conditions is discussed (for completeness) in Appendix \ref{app:BC}.

We illustrate the nonlinear adiabatic approach on numerical examples in the next Section.

\vspace{-0.3cm}
\section{Numerical Experiments: Validation of the Adiabatic Approach}
\label{sec:numerics}

In our proof-of-concept validation experiments we choose to experiment with
\begin{eqnarray}
\lambda(t)=\lambda_0(2+\cos(2\pi t/\tau))\cos(\pi t/\tau),\ G_0(t)=\mbox{const},
\label{exp_lambda0_G0}
\end{eqnarray}
thus testing the adiabatic approximations (UA and BA) in the regime where parameters of the exact solution, $\lambda$ and $G$,  may evolve on the time scale comparable to the time scale of the bare/exact solution,
and consider the following initial and boundary conditions:
\begin{eqnarray}
& p(0,x)&=p_{\mbox{\small UA}}(x;\lambda(0),G_0(0)),\label{init_exp}\\
& (pp):&\quad p(t,0)=p_{\mbox{\small UA}}(0;\lambda(t),G_0(t)),\nonumber\\ && p(t,L)=p_{\mbox{\small UA}}(L;\lambda(t),G_0(t)),
\label{bound_p_p}\\
&(p\phi):&\quad p(t,0)=p_{\mbox{\small UA}}(0;\lambda(t),G_0(t)),\nonumber\\ && p(t,L)=p_{\mbox{\small UA}}(L;\lambda(t),G_0(t)),
\label{bound_p_phi}
\end{eqnarray}
where $(pp)$ and $(p\phi)$ mark the two Boundary Condition (BC) cases (discussed in the main text and in the Appendix \ref{app:BC}, respectively).

For each BC setting we compare results of (a) numerical integration of Eqs.~(\ref{mass_balance_1},\ref{momentum_balance_reduced},\ref{exp_lambda0_G0},\ref{init_exp}) and Eq.~(\ref{bound_p_p}) or Eq.~(\ref{bound_p_phi}) [solid red in the Figures, the PDE integration is implemented in Mathematica];
(b) UA solution plus correction computed via numerical integration of the system of equations linearized around the UA solution, i.e. Eqs.~(\ref{adiabatic_p},\ref{adiabatic_phi},\ref{mass-balance-adiabatic},\ref{momentum-balance-adiabatic},\ref{exp_lambda0_G0},\ref{init_exp}) and Eq.~(\ref{bound_p_p}) or Eq.~(\ref{bound_p_phi}) [dashed red in the Figures, the PDE integration is implemented in Mathematica];
(c) UA solution, also correspondent to the case (b) with the second term on the lhs of Eq.~(\ref{adiabatic_p}) dropped [solid green in the Figures for the UA solution plus perturbation and dashed green for the UA solution only, where $\delta p$ is dropped];
(d) BA solution (see Appendix \ref{app:BA}) [solid blue in the Figures for the BA solution plus perturbation and dashed blue for the BA solution only, where respective $\delta p$ is dropped].

To simplify analysis, notations and visualizations we show results in the dimensionless/re-scaled units: time in the units of $T=3600s=1h$, $\tilde{t}=t/T$; distance in the units of $L=100km$, $\tilde{x}=x/L$; pressure in the units of $p_0=50 Bar=5\times 10^6 kg/(m s^2)$; and the flow in the units of $\phi_0= p_0 L/(c_s^2 T)\approx 1543 kg/m^2 s$.  In these re-scaled units Eqs.~(\ref{mass_balance_1},\ref{momentum_balance_reduced}) become
\begin{eqnarray}
\label{mass+momentum_balance_1_rescaled} &&
\partial_{\tilde{t}} \tilde{p}+\partial_{\tilde{x}} \tilde{\phi}=0,\quad
\partial_{\tilde{x}}\tilde{p}+\tilde{\alpha} \frac{\tilde{\phi}|\tilde{\phi}|}{2\tilde{p}}=0,
 \end{eqnarray}
where $\tilde{\alpha}=\alpha \phi_0^2 L/p_0^2=\alpha L^3/(c_s^4 T^2)\approx 8.57$.

Figures show results of our experiments for $\tilde{\lambda}_0=\lambda_0 L^{3/2}=0.05, \tilde{G}_0=G_0 L=0.3, \tau=2, \tilde{t}=5$ in the dimensionless units. We observe that
\begin{itemize}
\item The adiabatic approximation works reasonably well not only in the asymptotics, when parameters of the exact solution, $\lambda$ and $G$, are frozen (in time), but also in the borderline regimes when changes in the parameters occur on the time scale comparable with the one of the exact solution.
\item The nonlinear adiabatics introduced and discussed in this manuscript, generally, approximate exact results more accurately than the Linear Adiabatics.
\item Quality of the approximation is better for mass flows than for pressures.
\item Comparison of the two cases correspondent to the two types of boundary conditions suggests that the adiabatic approximation seems more accurate in the case when the two temporal profiles of pressure are fixed at the opposite ends of the pipe.
\item Beyond adiabatic linearization (keeping and not ignoring the second term on the lhs of Eqs.~(\ref{mass-balance-adiabatic})) improves nonlinear adiabatic approximation even more.
\end{itemize}

\begin{figure}[htb]
 \begin{center}
 \includegraphics[width = 0.5\textwidth]{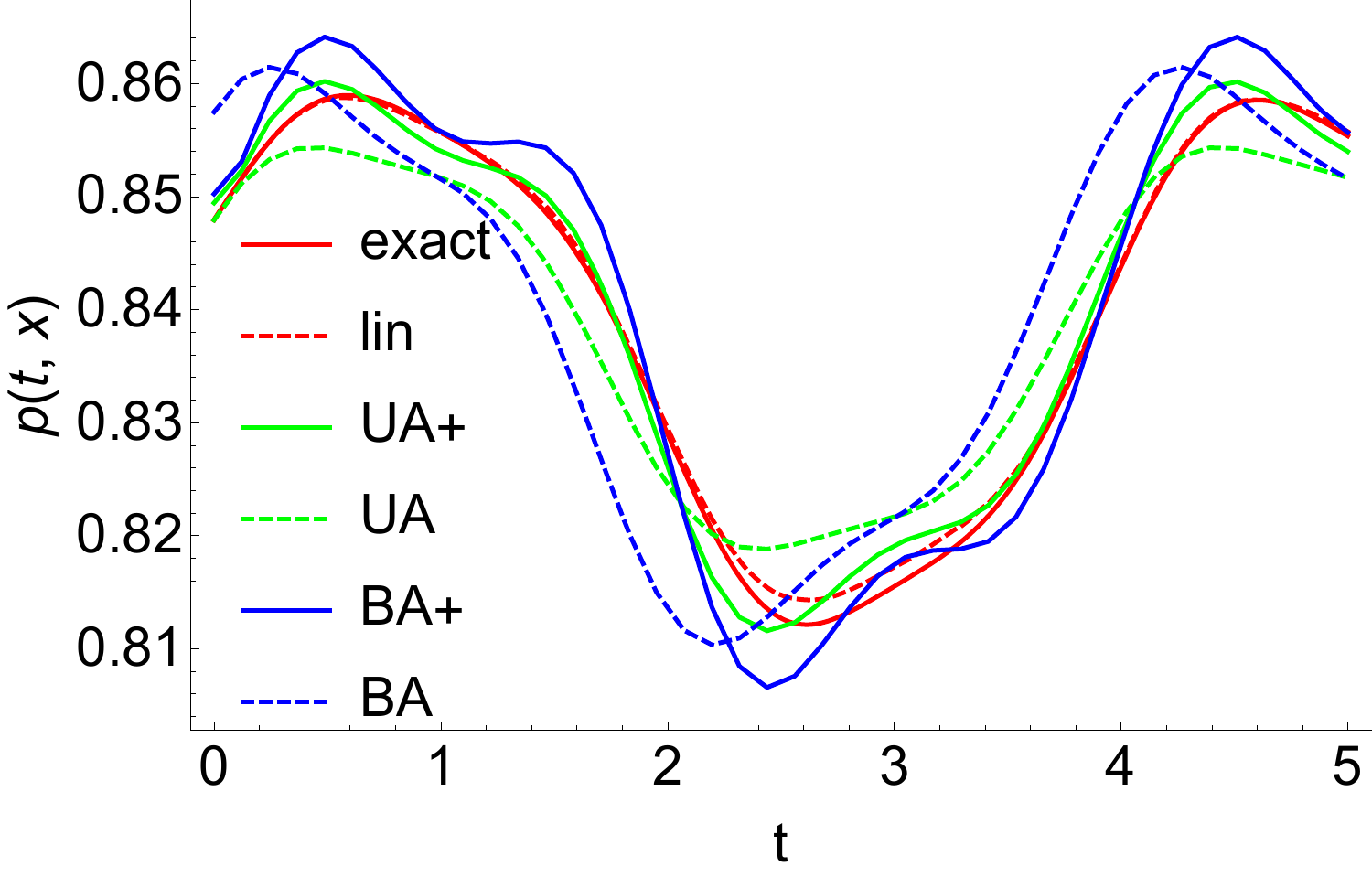}
 \includegraphics[width = 0.5\textwidth]{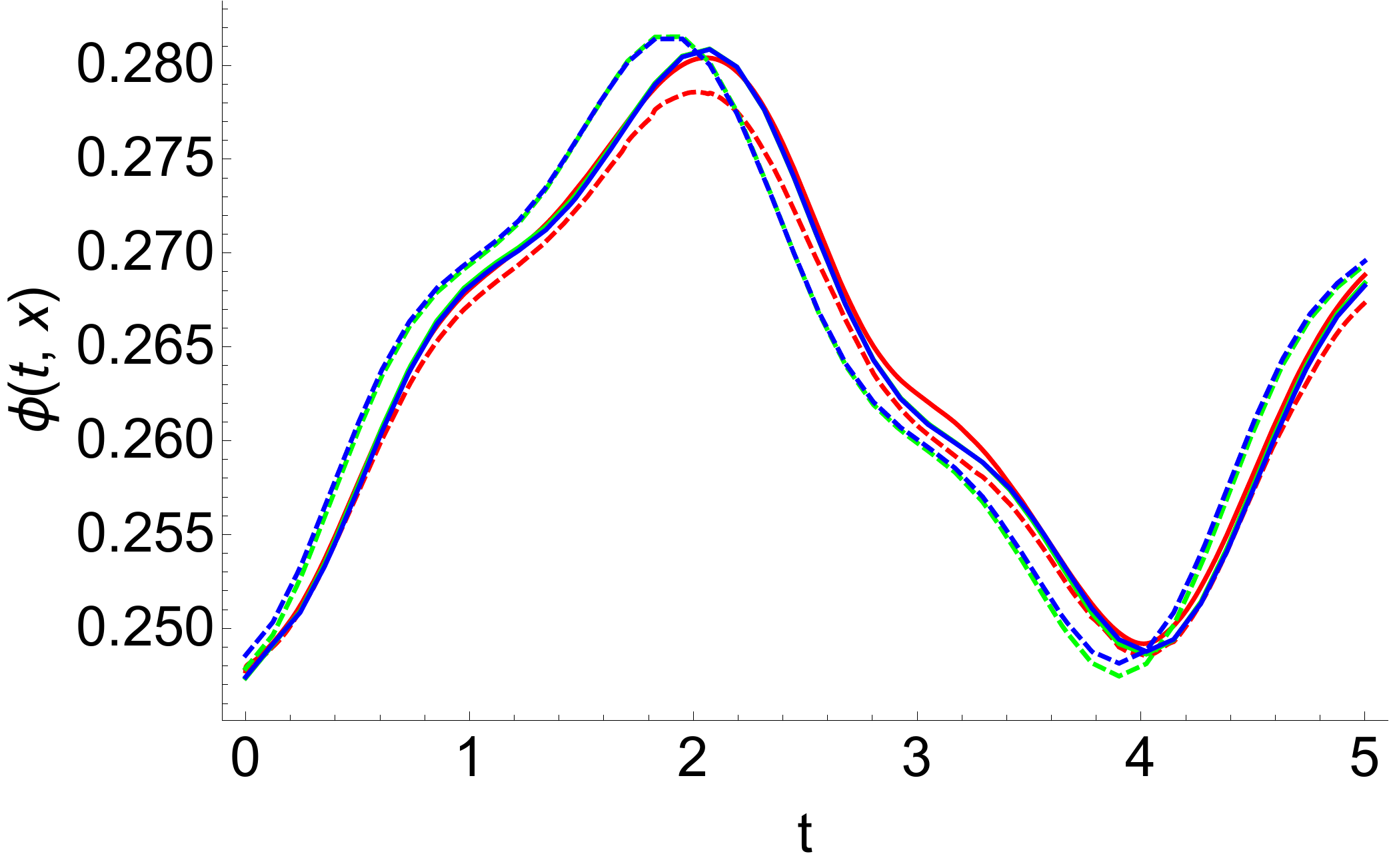}
 \end{center}
 \caption{\label{fig:PP1} Pressure profiles (top) and mass flow profiles (bottom) shown in the middle of the pipe for the case of the pressure boundary conditions fixed at the two ends of the pipe for direct numerical integration of Eqs.~(\ref{mass_balance_1},\ref{momentum_balance_reduced})
 [solid red - considered as the ground truth],
 UA solution plus linear correction computed via direct integration of Eqs.~(\ref{mass-balance-adiabatic},\ref{momentum-balance-adiabatic}) [dashed red], UA solution with perturbative correction [solid green], UA solution only [dashed green], BA solution with perturbative correction [solid blue], BA solution only [dashed blue].
 }
\end{figure}
\begin{figure}[htb]
 \begin{center}
 \includegraphics[width = 0.5\textwidth]{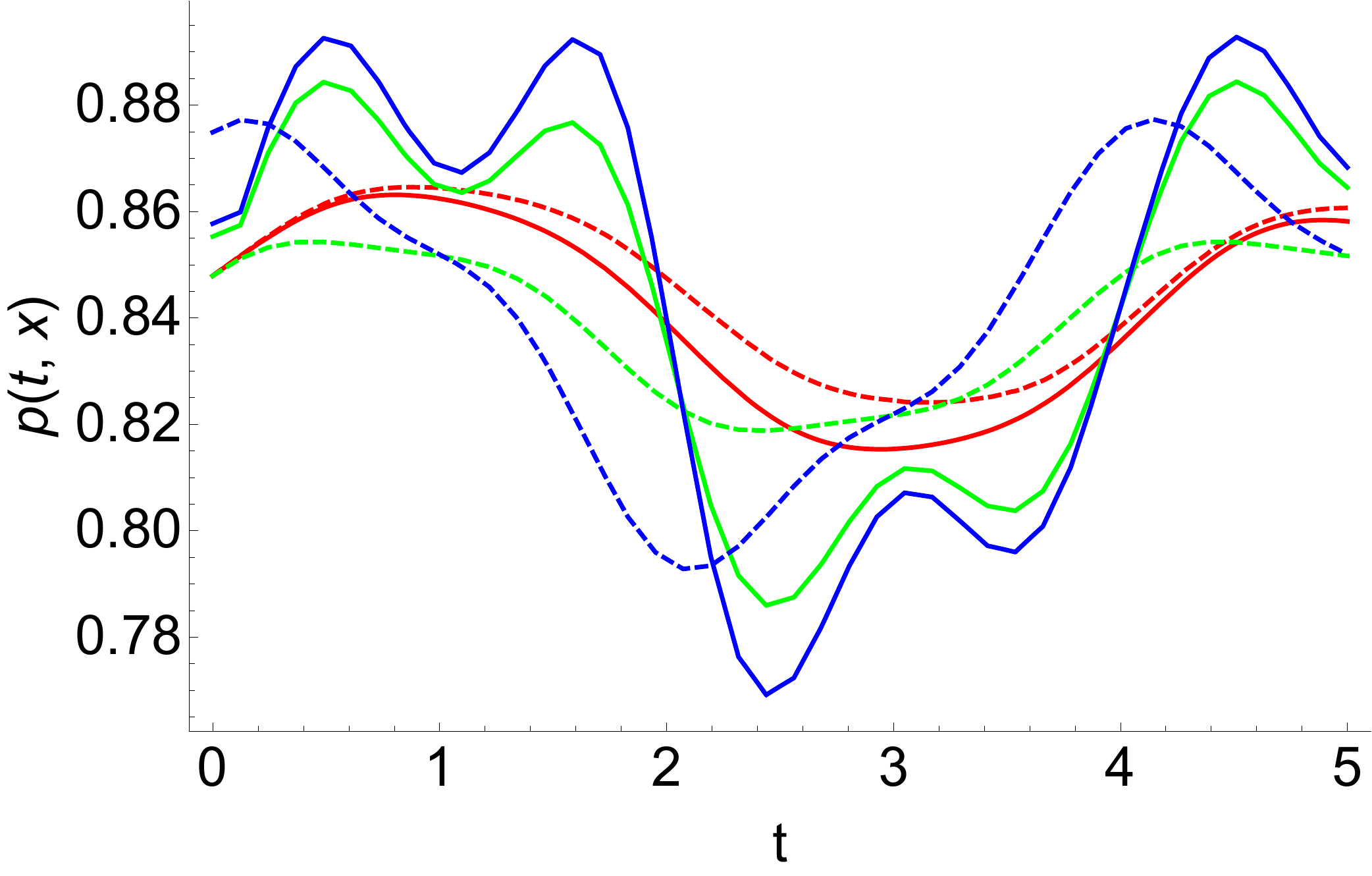}
 \includegraphics[width = 0.5\textwidth]{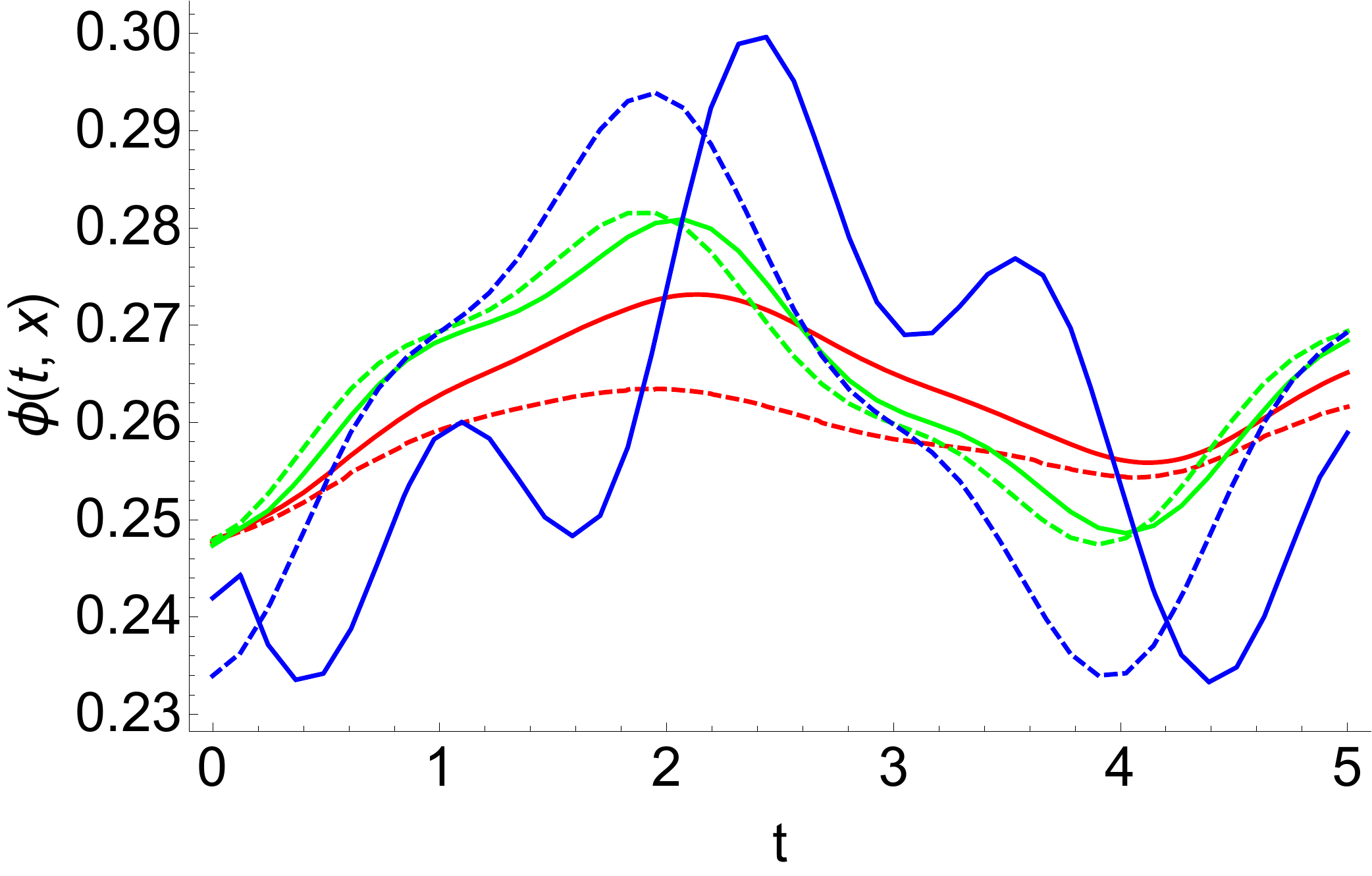}
 \end{center}
 \caption{\label{fig:PPhi1} Pressure profiles (top) and mass flow profiles (bottom) shown in the middle of the pipe for the case of the pressure boundary conditions fixed at the two ends of the pipe. Style and color-coding of lines is the same as in Fig.~(\ref{fig:PP1}).}
\end{figure}

\vspace{-0.3cm}
\section{Conclusions \& Path Forward}
\label{sec:consutions}

Our main conclusion is that the newly suggested Unbalanced Adiabatic solution outperforms the Balanced Adiabatic solution introduced originally in \cite{CBL2015} in the regimes where deviations from the balanced solutions occur sufficiently fast and are significant by amplitude. However, the improvement does not come for free -- even though the newly developed UA solution is more accurate (in terms of approximating the ground truth) it is also more complicated to find (stated implicitly via an inverse of an explicitly known transcendental function, while BA is stated explicitly in terms of elementary functions).

In the (relatively short term) future we plan to develop and test efficient computational schemes to model natural gas dynamics over extended networks, i.e. in the setting consisting of many elements (loads/generators, compressors, branching points, etc). It will be important to validate performance of UA and BA approximations in the realistic setting. To achieve an even better quality of approximation, we also plan to extend the approximation and account for linear corrections to the  solutions which are beyond the adiabatic approximation. Strategically, once the adiabatics (and beyond) are tested and validated,  we plan to utilize the approximations to boost solutions of multi-level optimization and control problems, e.g. of the type discussed in \cite{15ZCB,15ZDBC,15RC,15SP,15MVZHB,17ZRBCA}.

\vspace{-0.3cm}
\section{Acknowledgments}

The work of MC at LANL was supported in part by funding from the U.S. DOE/OE. The authors are thankful to a review for bringing to their attention the work on seemingly different problem (gas pipe line leak detection) \cite{PDE_leakage}, where a solution factorization method similar to our Eqs.~(\ref{p_pipe},\ref{phi_2_pipe}) was explored.

\appendices

\vspace{-0.2cm}
\section{Mixed (pressure and flow) Boundary Conditions}
\label{app:BC}

In the main part of the manuscript we describe the case when the pressure(s) are fixed at the two ends of the pipe. In this Appendix we discuss the mix case, when pressure is fixed only at one end of the pipe,  while fixed mass flow is controlled at the other end of the pipe. (Two clarifying remarks, on what would be other boundary conditions to consider, are in order. First, note that even though fixing pressure and flow at one side of the pipe would be formally/mathematically allowed, we do not believe that this case is physically/practically enforceable, thus excluding it from the consideration. Second, fixing two mass flow conditions at the two different ends of the pipe is not discussed either - now because we were not able to formulate a well-posed adiabatic approximation in this case.)

Consider pressure fixed at the ``incoming" (along the flow) side of the pipe and the mass flow is fixed at the outgoing side of the pipe. These conditions apply, e.g., when a consumer of gas, extracting a prescribed (and possibly time dependent) amount of the mass, is positioned at the outgoing side of the pipe. In this case, in addition to the boundary condition (\ref{p_boundary0}) at $x=0$ one also maintains the following boundary conditions at $x=L$:
\begin{eqnarray}
&& \phi(t,L)=p(t,0)\exp\left(-\int\limits_{0}^{L} dx' G(x';\lambda(t), G_0(t))\right)\nonumber\\ && \times\sqrt{\frac{2G(L;\lambda(t),G_0(t))}{\alpha}}.
\label{phi_boundaryL}
\end{eqnarray}
This setting assumes that $\delta p(t,0)=0$ and $\delta\phi(t,L)=0$. Then Eq.~(\ref{mass-balance-adiabatic-1}) is substituted by
\begin{eqnarray}
&&\delta \phi(t;x)=\int\limits_x^L dx'
\Biggl( \partial_{x'} \phi_{\mbox{\small UA}}(x';\lambda(t),G_0(t)) \\ && +c_s^{-2}\partial_t p_{\mbox{\small UA}}(x';\lambda(t),G_0(t))\Biggr),
\label{delta_phi2}
\end{eqnarray}
while (\ref{delta_p}) stay the same.

\vspace{-0.2cm}
\section{Balanced Adiabatic Approximation}
\label{app:BA}

The adiabatic approximation developed in this paper is a generalization of the Balanced Adiabatic (BA) approximation which was developed and discussed in details in \cite{CBL2015}. In this Appendix we briefly discuss/remind adaptation of the BA approximation to the cases with different boundary conditions (discussed in the main text and also in the preceding Appendix \ref{app:BC}).

The BA analogs of Eqs.~(\ref{p-base},\ref{phi-base}) are
\begin{eqnarray}
\label{p-BA}
&& p_{\mbox{\small BA}}(x;p_{in}(t),p_{out}(t))= \\
&&\nonumber \sqrt{(p_{in}(t))^2-\frac{x}{L}\left((p_{in}(t))^2-(p_{out}(t))^2\right)}\\ \label{phi-BA}
&& \hspace{-1cm}\phi_{\mbox{\small BA}}(p_{in}(t),p_{out}(t)) = \sqrt{\frac{(p_{in}(t))^2-(p_{out}(t))^2}{\alpha}}.
\end{eqnarray}
Then respective versions of Eqs.~(\ref{mass-balance-adiabatic-1},\ref{delta_phi_0},\ref{delta_p}) and all other follow up relations for the two cases of interest (pressures are fixed at the both sides of the pipe,
pressure and flow are fixed at the opposite sides of the pipe) are derived simply by substituting $p_{\mbox{\small UA}}, q_{\mbox{\small UA}}$ by $p_{\mbox{\small BA}}$, $q_{\mbox{\small BA}}$.

\bibliographystyle{IEEEtran}

\bibliography{cdc_paper}

\end{document}